\documentclass[]{aa}
\usepackage{graphicx}
\usepackage{txfonts}
\begin{document}
%\thesaurus{}
\title{Thermal and non-thermal energies of solar flares}
\author{P. Saint-Hilaire\inst{1,2} \and A.O. Benz\inst{1}}
\offprints{P. Saint-Hilaire}
\mail{shilaire@astro.phys.ethz.ch}
\institute{Institute of Astronomy, ETH Zentrum SEC, CH-8092 Zurich, Switzerland \and Paul Scherrer Institute, CH-5232 Villigen PSI, Switzerland}
\date{}
\titlerunning{Thermal and non-thermal energies of solar flares}
\authorrunning{P. Saint-Hilaire and A.O. Benz}
\abstract{The energy of the thermal flare plasma and the kinetic energy of the non-thermal electrons in 14 hard X-ray peaks from 9 medium-sized solar flares have been determined from RHESSI observations.
	The emissions have been carefully separated in the spectrum. 
	The turnover or cutoff in the low-energy distribution of electrons has been studied by simulation and fitting, yielding a reliable lower limit to the non-thermal energy.	
	It remains the largest contribution to the error budget.
	Other effects, such as albedo, non-uniform target ionization, hot target, and cross-sections on the spectrum have been studied. 
	The errors of the thermal energy are about equally as large.
	They are due to the estimate of the flare volume, the assumption of the filling factor, and energy losses. 
	Within a flare, the non-thermal/thermal ratio increases with accumulation time, as expected from loss of thermal energy due to radiative cooling or heat conduction. 
	Our analysis suggests that the thermal and non-thermal energies are of the same magnitude. 
	This surprising result may be interpreted by an efficient conversion of non-thermal energy to hot flare plasma.
	}
\date{Received ... / Accepted ...}

\maketitle
%------------------------------------------------------------------------------
\section{Introduction}
In the standard model of solar flares, a major part of the energy is first released into energetic, non-thermal electrons and possibly ions. 
These particles, guided by magnetic field lines, may be lost in interplanetary space, but also precipitate into the lower corona or upper chromosphere where they loose their energy by Coulomb collisions with the denser medium. 
This energy is believed to heat up the ambient plasma to tens of millions of degrees, which will rise and fill the coronal loop. 
An interesting question is thus the relation between non-thermal and thermal energies. 
The difference between them may indicate conversion losses and/or other forms of primary energy release. 

How does total kinetic energy of electrons that precipitated compare to the thermal energy of the plasma? 
In the pre-RHESSI era and in RHESSI first results papers, this issue had been addressed (see e.g. de Jager et al. \cite{deJager1989}; Saint-Hilaire and Benz 2002), with the result that the kinetic energy was often reported to be up to an order of magnitude or more than the observable thermal energy. 
Others (Gan et al. 2001) have managed to conclude just the opposite. 

The hot thermal plasma emits soft X-rays. 
They differ in the spectrum from the X-rays emitted by the non-thermal electron bremsstrahlung at higher energies. 
Soft and hard X-rays can thus be used to determine the thermal and non-thermal energies, respectively. 
In practice, however, many uncertainties limit the precision of the energy determination. 
Unfortunately, the two emissions are usually cospatial and overlap in photon energies in the range from about 10 to 25 keV. 
This is furthermore the range of the emission of the non-thermal photons that carry the information on the dominant lowest-energy part of non-thermal energies. 
The distinction between the two forms of energy requires a high spectral resolution in the critical range. 

An accurate derivation of non-thermal energy from the spectrum is not trivial. 
In previous work the observed hard X-ray spectrum was converted into electron energies assuming that they impinge onto a thick target. 
The energies were integrated starting at some assumed lower limit or at the crossover between thermal and non-thermal emissions in the observed photon spectrum. 
As the derived electron energy distribution is a power law with large index, the result depends much on the lower bound of integration. 
In addition, the photon spectrum is influenced by several effects. 
The non-thermal bremsstrahlung of a coronal source is reflected by the dense layers below, and thus appears to be brighter (Bai \& Ramaty 1978; Alexander \& Brown 2002).
Precipitating electrons above a certain energy penetrate into the chromosphere where they lose energy by collisions with neutrals and are more efficient in bremsstrahlung radiation than in the completely ionized corona (Brown, 1973; Kontar et al. 2002). 
Additionally, the Sun is not a simple cold, thick target (Emslie 2003).
Finally, the various approximations for electron cross-section used in the literature yield different values in particular at relativistic particle energies. 

On the other hand, the accurate determination of the thermal energy also poses problems. 
Thermal energies are best estimated from the thermal bremsstrahlung spectrum, yielding the plasma temperature and the emission measure. 
The volume of the thermal plasma must be estimated from the size of the source and an assumption on the filling factor. 
The ambient plasma cools down by either heat conduction or radiative cooling (i.e. thermal bremsstrahlung of an optically thin plasma) (Porter and Klimchuck 1995; Aschwanden et al. 2001). 
Thus, thermal energy is lost as it is being measured. 
A reliable determination of the errors is therefore as important as the final ratio between the two forms of energy.

In this paper, we determine energy budgets for several flares observed by the Reuven Ramaty High Energy Solar Spectroscopic Imager (RHESSI) (Lin et al. 2002). 
The imaging capabilities, high spectral resolution and broad photon energy coverage of RHESSI make this instrument ideal to determine the two energies with much higher precision and study their relation. 
Medium-sized flares (upper C and lower M class) have been selected to avoid photon pile-up, complicated source structure and other conundrums of larger flares. With a careful RHESSI analysis, some -but not all- of the uncertainties or assumptions used by previous authors may be removed, and more accurate results obtained.

This paper will start with a section discussing some already established facts of solar flare bremsstrahlung emissions and the various high-energy electron cross-sections available in the literature. The next section will deal with the method proposed to determine flare energies. 
Observational results from RHESSI will then be presented, and the ratios of {\it cumulative non-thermal energy} over {\it thermal energy increase} (during the same time interval) will be examined.

%------------------------------------------------------------------------------
\section{Basic theory}
	\subsection{Thick-target bremsstrahlung emission}
		An initial (injection) electron beam with distribution $F_0(E_0)$ electrons s$^{-1}$ keV$^{-1}$ passing through a dense --although still optically thin-- plasma emits
		HXR radiation according to the following formula, for a {\it thick target} (Brown 1971):
		\begin{equation}\label{Ithick}
			I_{thick} =\frac{1}{4\pi D^2} \int_{E_0=\epsilon}^{\infty} F_0(E_0) \int_{E=\epsilon}^{E_0}  \frac{Q_B(E,\epsilon)}{E \, Q_c(E)} \,\, dE \,\, dE_0 ,
		\end{equation}
		where $I_{thick}(\epsilon)$ is the observed photon spectrum seen at distance $D$ (assumed to be 1 AU in the following) from the site of emission, in photons s$^{-1}$ cm$^{-2}$ keV$^{-1}$.
		Isotropic emission is assumed throughout this work.
		$Q_B(E,\epsilon)$ is the bremsstrahlung differential cross-section, for an electron of energy $E$ emitting a photon of energy $\epsilon$.
		$Q_c(E)$ is the energy-loss cross-section due to Coulomb collisions with the ambient plasma.				

%		In the thick-target model, the electrons are assumed to loose all their energy due to Coulomb collisions (thereby heating the local medium) within the spatial resolution of the instrument.
%		Energy losses due to bremsstrahlung are neglected (they start to be competitive for electrons above $\sim$300 MeV: those energies will not interest us in this work).
%		The inner integral corresponds to the photon spectrum emitted by an electron of initial energy $E_0$, being slowed down to energy $\epsilon$.
%		Photons are assumed to reach the observer unimpeded.

Using non-relativistic cross-sections, Brown (1971) has demonstrated that there exist an analytical relationship between $I_{thick}(\epsilon)$ and $F_0(E_0)$, if $F_0(E_0)$ is a power-law.
If $F_0(E_0)=A_eE^{-\delta}$, then $I_{thick}(\epsilon)=A_{\epsilon} \cdot \epsilon^{-\gamma}$, with:
		\begin{eqnarray}
			\label{Ithick2}
			A_{\epsilon} 		&=&	\frac{A_e}{4 \pi D^2} \overline{Z^2} \frac{\kappa_{BH}}{K} \frac{B(\delta-2,1/2)}{(\delta-1)(\delta-2)}	\\
			\label{Ithick2B}			
			\gamma			&=&	\delta-1												
		\end{eqnarray}
		where $A_e$ is in electrons s$^{-1}$ keV$^{-1}$, $\overline{Z^2}$ is 1.44 for typical coronal abundances, $\kappa_{BH}=\frac{8}{3} \, \alpha \, {r_e}^2 \, m_ec^2 \, = \, 7.9 \times 10^{-25}$ cm$^2$ keV, $K=2\pi e^4\Lambda \, = \, 2.6 \times 10^{-18}$ cm$^2$ keV$^{-2}$ for a fully ionized plasma, and $B$ is the beta function [$B(a,b)=\frac{\Gamma(a)\Gamma(b)}{\Gamma(a+b)}$] (Tandberg-Hanssen and Emslie 1988).
		Numerically,
		\begin{equation}\label{IthickNum}
			I_{thick}(\epsilon) = 1.51 \times 10^{-34} \frac{B(\delta-2,1/2)}{(\delta-1)(\delta-2)} \,A_e\,\epsilon^{1-\delta} \, ,
		\end{equation}

\begin{figure}
		\centering
		\includegraphics[width=8.8cm]{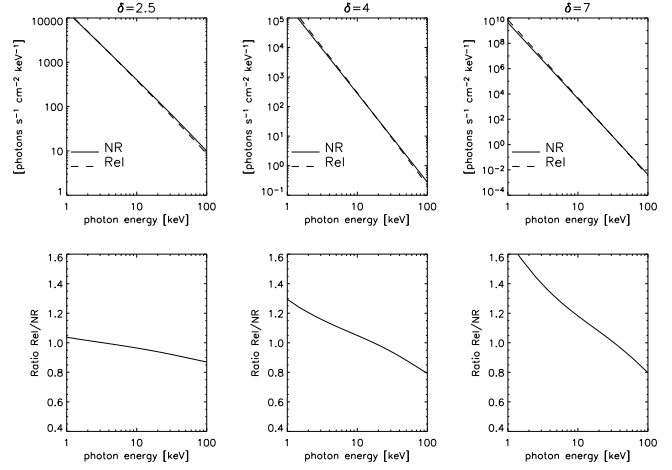}	%figs/NRvsRel_cross_Eto20.ps
		\caption{Synthetic photon spectra generated from perfect injection electron power-law spectra of varying spectral index $\delta$.
			 {\it Top row:} The {\it solid line} were computed using the same non-relativistic cross-sections as Brown (1971).
			 The {\it dashed line} were computed with the Haug (1997) cross-section and the Bethe-Bloch formula for energy loss.
			 The {\it bottom row} displays the ratio between the two.
			 }			 
		\label{NRvsRelCross}
		\end{figure}

\begin{figure}
		\centering
		\includegraphics[width=8.8cm]{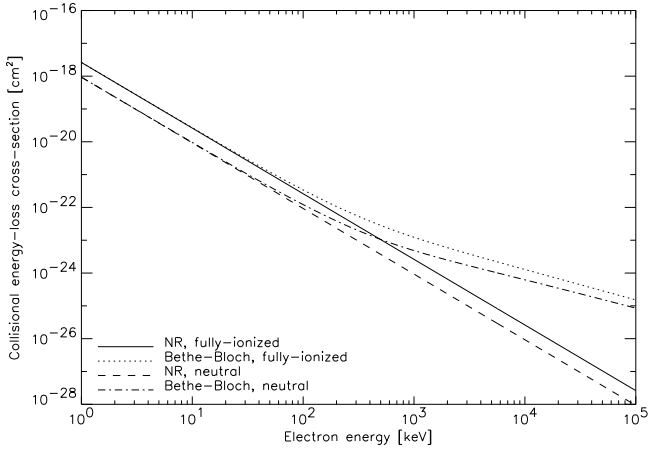}
		\caption{Energy-loss cross-section, for the usual non-relativistic (NR) case and for the Bethe-Bloch formula.
			Notice also the slight dependence of the spectral index at relativistic energies with the ionization level.
			The spectral index at high energies is $\sim 0.93$ for a fully-ionized plasma, and $\sim 0.87$ for a neutral medium.}
		\label{bbcross}
		\end{figure}

		Figure \ref{NRvsRelCross}  shows the difference when using more accurate relativistic cross-sections:
		The Haug (Haug 1997) differential bremsstrahlung cross-section and the full quantum relativistic Bethe-Bloch formula (Longair 1992) for energy loss:
		\begin{equation}\label{Bethe-Bloch}	
			-\left(\frac{dE}{dx}\right)_{coll}=\frac{4 \pi e^4}{m_e v^2} n_e \left( \Lambda + ln(\gamma^2) -\beta^2 \right) \,\,\, ,
		\end{equation}		
		where $\Lambda$ is the usual non-relativistic Coulomb logarithm, $\gamma$ the Lorentz factor, and $\beta=\frac{v}{c}$.
In its non-relativistic (NR) limit, the Bethe-Bloch formula has the $\sim \frac{1}{E^2}$ dependence, up to electron kinetic energies nearing the electron rest mass energy, after which the dependence is $\sim\frac{1}{E}$ (Fig. \ref{bbcross}).
We note here that the Bethe-Bloch cross-section is very close to what is used in RHESSI software.

%Hence, at high energies, Coulomb losses decrease less rapidly than with the NR approximation (Fig. \ref{bbcross}), leading to a softening of the emitted thick-target photon spectra, as shown in Fig. \ref{bethebloch}. 
%At non-relativistic energies, the effect is only important for hard electron distributions. 
%Using the NR cross-section thus underestimates the power-law index of the electron distribution even at NR photon energies. 

Other effects, such as photon back-scattering on the photosphere (Bai \& Ramaty 1978; Alexander \& Brown 2002), non-uniform target ionization (Brown 1973; Kontar et al. 2002), or a hot target (Emslie 2003) need to be considered.

		The non-thermal kinetic power of the injected electron beam may be computed:
		\begin{equation}
			P = \int_{?}^{\infty} E \cdot F_0(E) \,\,\, dE
		\end{equation}
		The introduction of some kind of a cutoff at low energies is necessary by the fact that the integral diverges at zero energy. Assuming that $F_0(E_0)$ remains a power-law below the thermal energy in the acceleration region is not physical.
		In the past, a sharp low-energy cutoff in $F_0(E_0)$:
		\begin{equation}\label{EQ_cutoff}
			 F_0(E_0) = \left\{ \begin{array}{ll}
				A_e E_0^{-\delta} & \mbox{for $E_0>E_{co}$} \\
				0 		& \mbox{for $E_0<E_{co}$} 
						\end{array}
					\right. 
		\end{equation}
		was often assumed.
		This situation seems physically not realistic as such a configuration leads to plasma instabilities. 
		Such instabilities have growth rates typically of the order of the local plasma frequency, i.e. orders of magnitude shorter than the propagation time of the beam within the acceleration region. 
		The (flat) turnover model seems physically closer to reality:
		\begin{equation}\label{EQ_turnover}
			 F_0(E_0) = \left\{ \begin{array}{ll}
				A_e E_0^{-\delta} 	& \mbox{for $E_0>E_{to}$} \\

				A_e E_{to}^{-\delta}	& \mbox{for $E_0<E_{to}$} 
						\end{array}
					\right. 
		\end{equation}		
We note, however, that recent simulations of stochastic electron acceleration, such as by Petrosian \& Liu (2004) predict an energy distribution in the acceleration region still increasing below the turnover energy.
Coulomb losses tend to linearize the low-energy part of an electron energy distribution.
From their resulting photon spectra, such a linearization below the turnover would be extremely difficult to distinguish from a flat turnover, as it effects the least energetic photons, where the thermal emission usually dominates (see also the next paragraph).
Thus, the flat turnover model may not be the true distribution, but yields rather an upper limit to the total non-thermal energy estimate. 
%As discussed in Appendix A, collisional energy losses during propagation in the corona tend to impart such a shape to the electron spectrum before it reaches the denser lower corona or upper chromosphere and emits most of the bremsstrahlung. Thus, these low-energy electrons mainly heat the corona and do not contribute to the evaporation.

		\begin{figure}
		\centering
		\includegraphics[width=8.8cm]{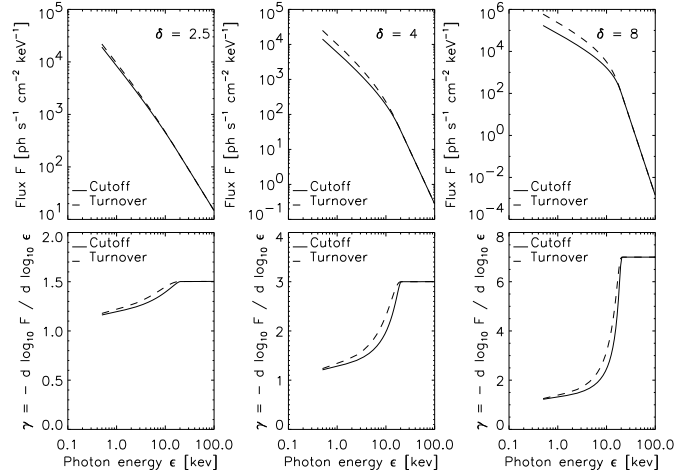}
		\caption{{\it Top:} Photon spectra and their spectral indices ({\it bottom}), computed from different injection electron power-laws of spectral indices $\delta$, using the Brown (1971) cross-sections. 
			The electron spectra had all a 20 keV cutoff {\it (solid line)} or turnover {\it dashed line} energy.}
		\label{cutoff_turnover_deriv}
		\end{figure}

The photon spectra of the different model distributions of electrons in a beam impinging on a thick target have been calculated using the Brown (1971) model (i.e. using the non-relativistic Bethe-Heitler bremsstrahlung cross-section, and non-relativistic collisional losses). 
Both turnover and cutoff model lead to a photon spectrum that is rounded off at low energies (see Fig. \ref{cutoff_turnover_deriv}). 
The spectral index of the turnover model is slightly larger than for the cutoff model. 
Note that it is not a constant value of $\sim$1.5 as is sometimes assumed. 
Both tend asymptotically towards $\sim$1.15 for all $\delta$ as the photon energy approaches zero. 
The usually observed superposition of a thermal component (or a full differential emission measure distribution) to the power-law spectrum further makes an observational distinction in the spectrum exceedingly difficult. 
%As the cutoff model has been widely used in the literature and even seems to be supported by recent RHESSI results, both have been examined in the following.

		If $F_0(E_0)$ has a cutoff with the shape defined in Eq. (\ref{EQ_cutoff}), the non-thermal kinetic power contained in the beam of electrons is given by:
		\begin{equation}\label{nonthermal_power_cutoff}
			%P = 6.62 \times 10^{33} \,\,\, \frac{\gamma}{B(\gamma-1,1/2)} Eco^{1-\gamma} \,\,\, [erg/s]
			P_{cutoff} = \frac{A_e}{\delta-2} E_{co}^{-\delta+2}\ \ .
		\end{equation}
		On the other hand, if $F_0(E_0)$ has the form of a turnover as described in Eq. (\ref{EQ_turnover}),
		\begin{equation}\label{nonthermal_power_turnover}
			%P = 6.62 \times 10^{33} \,\,\, \frac{\gamma (\gamma-1)}{B(\gamma-1,1/2)} Eco^{1-\gamma} \left( \frac{1}{\gamma-1} + \frac{1}{2}\right) \,\,\, [erg/s]
			P_{turnover} = \frac{A_e}{\delta-2} E_{to}^{-\delta+2} \left(1 + \frac{\delta-2}{2} \right)\ \ .\end{equation}

		\begin{figure}
		\centering
		\includegraphics[width=8.8cm]{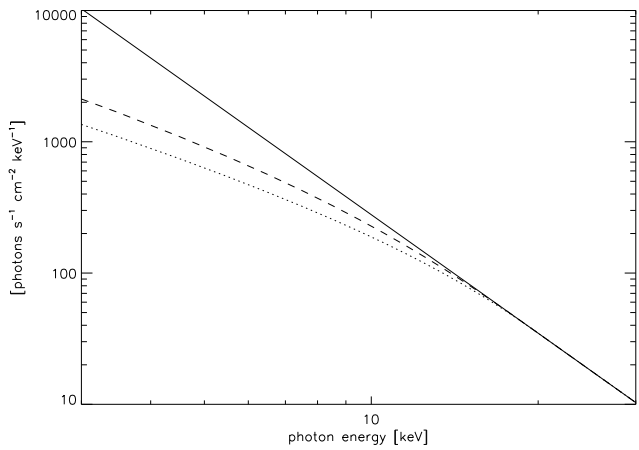}
		\caption{Photon spectrum observed at 1 AU, produced by an electron power-law distribution of spectral index $\delta$=4, and a 50 keV electron flux of $A_{50}=1.295\times10^{33}$ $electrons$ $s^{-1}$ $keV^{-1}$.
			{\it Solid line}: No cutoff. {\it Dotted line}: Cutoff at 20 keV. {\it Dashed line}: Turnover at 20 keV.}
		\label{blowup_none_cutoff_turnover}
		\end{figure}

\begin{figure}
		\centering
		\includegraphics[width=8.8cm]{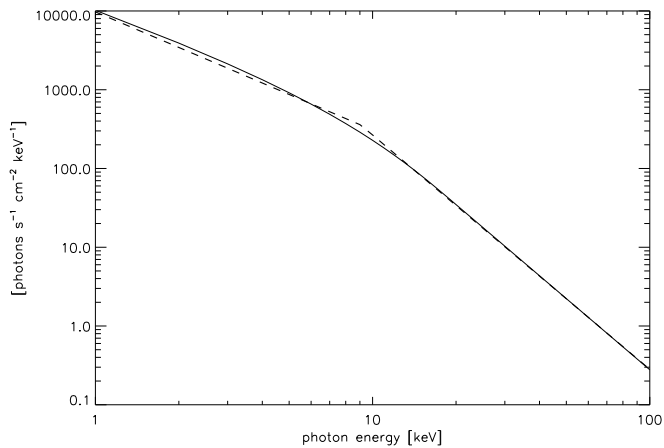}
		\caption{Fitting a photon spectrum with power-laws.
			A synthetic photon spectra (solid line) generated from an electron power-law with $E_{to}$=20 keV and $\delta$=4 (dashed curve in Fig. \ref{blowup_none_cutoff_turnover}) was fitted with a broken power-law (dashed line).
			The power-law left of the break energy $\epsilon_{to}$ had a fixed spectral index of 1.5.			
			For this example, fitting yields $\epsilon_{to}$=9 keV and a photon spectral index right of this break of $\gamma$=2.97 ($\approx\delta$-1).}
			%and the scaling factor is 1.01 times the real one.
		\label{fitting_bpow}
		\end{figure}

\begin{figure}
		\centering
		\includegraphics[width=8.8cm]{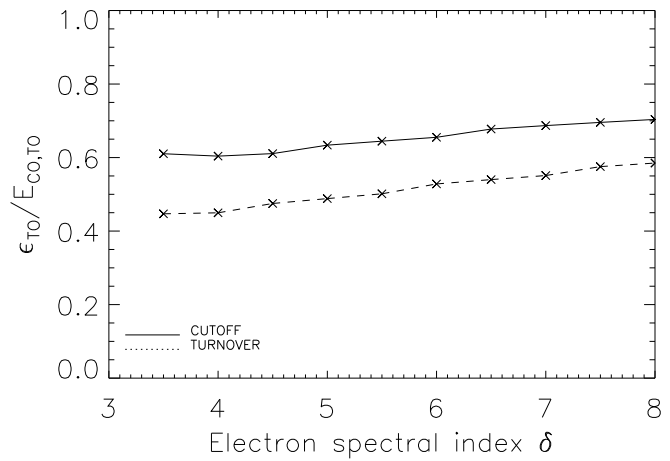}
		\caption{The ratio of the fitted turnover energy in the photon spectrum, $\epsilon_{to}$, to the cutoff resp. turnover energy in the electron beam distribution, as a function of the electron spectral index right of $E_{co,to}$. }
		\label{EphTO_Ecto_vs_delta}
		\end{figure}

Figure \ref{blowup_none_cutoff_turnover} is an enlargement of the turnover region in the photon spectrum. 
Note that the apparent photon turnover energy in Fig. \ref{blowup_none_cutoff_turnover} is below the electron cutoff/turnover energy (20 keV).
This has been further investigated in Figs. \ref{fitting_bpow} and \ref{EphTO_Ecto_vs_delta}. Synthetic photon spectra were produced from electron distributions with different power-law indices $\delta$, using the Brown (1971) approximation for easy comparison with analytical results.
The spectra were then fitted with a double power-law of spectral index 1.5 below the break. %, and spectral index $\gamma$ above it. 
The intersection of the two power-laws defines a photon turnover energy $\epsilon_{to}$ which is far below the electron cutoff energy  $E_{co}$ or turnover energy $E_{to}$.
Fig. \ref{EphTO_Ecto_vs_delta} displays the ratio of the photon turnover energy $\epsilon_{to}$ to either $E_{co}$ or $E_{to}$.
These ratios are smaller for the cutoff model than for the turnover model. 
If the photon turnover energy $\epsilon_{to}$ is used instead of $E_{co}$ or $E_{to}$, the derived non-thermal power in the electron beam may be overestimated by more than an order of magnitude.

\section{Known and unknown errors in computing non-thermal energies.}

So far, the emphasis has been on the difference due to the use of accurate cross-sections (both bremsstrahlung and Coulomb losses) or their NR approximations.
In this section, other approximations and factors are presented that also influence the photon spectrum and thus are possible sources of error for the derived electron energy distribution. 
Other effects on the photon spectrum are non-uniform target ionization (Brown 1973) and the albedo effect (Compton back-scattering on the photosphere, Bai and Ramaty, 1978). 
They complicate the observed photon spectra by the simple fact that the amplitude of their effects varies with the energy.
Kontar et al. (2002) for non-uniform target ionization and Alexander \& Brown (2002) for the albedo effect have provided corrections. 
We have used their formulas to compute correction factors in the numerical examples given below.

Finally, the possible presence of some high-energy cutoff (or break) in the injected electron distribution affects the photon spectrum at lower energies. The effect has been simulated numerically. 
A sharp electron high-energy cutoff at energy $E_h$, would lead to a noticeable deviation from power-law behavior starting already at photon energies above $\sim\frac{E_h}{3}$, where a pronounced rollover in the spectrum should occur. 
This was not observed in our selection of (mostly M-class) flares, at least below 35 keV, the upper limit of the fitting interval used in our data analysis.

			\begin{table}				
			\centering
			\begin{tabular}{l c c c c}
				\hline \hline

				Type of correction	& Quantity	& $\delta$=3	& $\delta$=5	& $\delta$=7	\\ 
				\hline
				Haug, Bethe-Bloch	& $\Delta\gamma$&	+0.1	&	+0.1	&	+0.1	\\
							& $f_{50}$	&	0.89	&	0.91	&	0.94	\\
							& $f_{CO}$	&	1.07	&	1.16	&	1.23	\\
							& $f_{TO}$	&	1.02	&	1.14	&	1.22	\\							
				\hline
				Haug, Bethe-Bloch,	& $\Delta\gamma$&	-0.2	&	0.0	&	+0.1	\\
				Albedo			& $f_{50}$	&	1.54	&	1.36	&	1.25	\\
							& $f_{CO}$	&	0.96	&	1.285	&	1.40	\\
							& $f_{TO}$	&	1.09	&	1.29	&	1.40	\\							
				\hline
				Haug, Bethe-Bloch,	& $\Delta\gamma$&	+2.1	&	+1.2	&	+0.7	\\
				High-E cutoff \@ 50 keV	& $f_{50}$	&	0.02	&	0.2	&	0.47	\\
							& $f_{CO}$	&	2.64	&	2.20	&	1.71	\\
							& $f_{TO}$	&	1.45	&	1.95	&	1.65	\\							
				\hline
				Haug, Bethe-Bloch,	& $\Delta\gamma$&	+0.35	&	+0.1	&	+0.1	\\
				High-E cutoff \@ 300 keV& $f_{50}$	&	0.50	&	0.90	&	0.94	\\
							& $f_{CO}$	&	1.23	&	1.17	&	1.23	\\
							& $f_{TO}$	&	1.02	&	1.15	&	1.22	\\							
				\hline
				Haug, Bethe-Bloch,	& $\Delta\gamma$&	0.0	&	0.90	&	0.0	\\
				$E_*$ = 5 keV		& $f_{50}$	&	2.54	&	2.67	&	2.84	\\
							& $f_{CO}$	&	2.81	&	2.74	&	2.63	\\
							& $f_{TO}$	&	2.74	&	2.74	&	2.64	\\							
				\hline
				Haug, Bethe-Bloch,	& $\Delta\gamma$&	-0.2	&	-0.4	&	-0.3	\\
				$E_*$ = 25 keV		& $f_{50}$	&	2.56	&	2.30	&	1.86	\\
							& $f_{CO}$	&	1.51	&	1.02	&	1.01	\\
							& $f_{TO}$	&	1.74	&	1.09	&	1.03	\\							
				\hline
				Haug, Bethe-Bloch,	& $\Delta\gamma$&	-0.2	&	0.0	&	+0.1	\\
				$E_*$ = 100 keV		& $f_{50}$	&	1.74	&	1.02	&	0.96	\\
							& $f_{CO}$	&	0.95	&	1.10	&	1.23	\\
							& $f_{TO}$	&	1.12	&	1.10	&	1.22	\\							
				\hline
				Haug, Bethe-Bloch,	& $\Delta\gamma$&	-0.4	&	-0.5	&	-0.4	\\
				$E_*$ = 25 keV,		& $f_{50}$	&	4.42	&	3.45	&	2.46	\\
				Albedo			& $f_{CO}$	&	1.24	&	1.12	&	1.15	\\
							& $f_{TO}$	&	1.84	&	1.22	&	1.18	\\							
				\hline
			\end{tabular}
			\caption {Differences in photon spectral indices and normalization factor (flux at 50 keV), as well as computed non-thermal power (cutoff and turnover cases), between reality and assuming the Brown (1971) model.
				See text for details.
				$E_*$, in the context of the non-uniform target ionization model, is the minimum initial energy an electron requires to reach the neutral parts of the chromosphere (same as in Brown 1973; Kontar et al. 2002).}
			\label{errors_example}
			\end{table}

\subsection{Some numerical examples}

How far from the truth are we if, from an observed photon power-law in the 10-35 keV band, we derive its electron power-law energy distribution using the Brown (1971) thick-target model (i.e. perfect power-law at all energies, non-relativistic cross-sections, cold target, no albedo, uniform 100\% ionization)? 
To estimate the effect of the different correction factors, we have computed some numerical examples. 
Synthetic photon spectra were computed from ideal power-law electron distributions with index $\delta$, using the relativistic cross-sections, and other effects (high-energy cutoffs, ionization, and albedo: See Table \ref{errors_example}).
These synthetic photon spectra were fitted in the 10-35 keV band (1-keV bins) with photon power-laws.
Photon spectral indices, $\gamma$, and fluxes at 50 keV, $F_{50}$, were determined from the fits. 
Using the Brown (1971) model, approximations to the original electron spectral indices (Eq. \ref{Ithick2B}) and normalization factors (Eq. \ref{Ithick2}) can then be determined, from which non-thermal powers can be computed using Eqs. (\ref{nonthermal_power_cutoff}) or (\ref{nonthermal_power_turnover}) (a 10 keV cutoff or turnover energy was assumed here: This arbitrary value does not greatly change the generality of the problem).
We call these quantities $\delta_{approx}$, $A_{e,approx}$, $P_{cutoff}^{approx}$, and $P_{turnover}^{approx}$.
$\Delta\gamma$ (Table \ref{errors_example}) is the difference between the spectral index $\gamma$ of the synthetic spectrum and $\delta-1$, the photon spectral index that would have been obtained with the Brown (1971) model (Eq. \ref{Ithick2B}).
Similarly, $f_{50}$ is the ratio of the photon fluxes at 50 keV of the synthetic spectra with those derived theoretically using the Brown (1971) model (Eq. \ref{Ithick2}).
If $P_{cutoff}^{real}$ and $P_{turnover}^{real}$ are the real non-thermal powers (derived from Eqs (\ref{nonthermal_power_cutoff}) and (\ref{nonthermal_power_turnover})), the factors $f_{CO}$ and $f_{TO}$:
		\begin{equation}\label{f_CO}
			f_{CO}=\frac{P_{cutoff}^{real}}{P_{cutoff}^{approx}}
		\end{equation}
		\begin{equation}\label{f_TO}
			f_{TO}=\frac{P_{turnover}^{real}}{P_{turnover}^{approx}}
		\end{equation}
are the corrections that must be applied to the rough $P_{cutoff,turnover}^{approx}$ estimates.
Table \ref{errors_example} indicates that the $\Delta\gamma=\gamma-(\delta-1)=\delta_{approx}-\delta$ is usually slightly positive, i.e. applying the Brown (1971) model to observed photon power-law spectra to determine the original electron power-law spectral index generally slightly underestimates that electron spectral index.
%Thus, applying no corrections and NR physics typically yields an electron energy distribution index $\delta$ that is too small. The uncorrected version overestimates the flux at 50 keV and consequently underestimates non-thermal energies by up to 20-30\%.
In the 10-35 keV range, for usually observed spectral hardnesses ($\delta >4$), both corrections to the non-thermal power for relativistic effects and albedo are of the same importance, about 15-20\%.
The correction due to non-uniform target ionization is usually more important, particularly for low $\delta$ and low $E_*$ ($E_*$ is the initial energy that electrons need in order to penetrate into the unionized chromosphere).
Some of the effects, when combined, might cancel each other out, and all depend on the spectral index $\delta$: General error estimates from each effect (or the sum of them) can only be done for a certain energy band of observation and if an approximate spectral index is known.
The $f_{TO,CO}$ corrections on non-thermal power exceed 50\% only in extreme cases. 
The biggest uncertainty comes by far from the low-energy cutoff or turnover energy: the non-thermal power going as $E_{co,to}^{-\delta+2}$. 
The addition of albedo and/or non-uniform target ionization effects combined with the fact that spectral fitting is somewhat model-dependant may displace this $E_{co,to}$.

%\vfill\eject

\subsection{Finding the low-energy cutoff or turnover}

		\begin{figure}
		\centering
		\includegraphics[width=8.8cm]{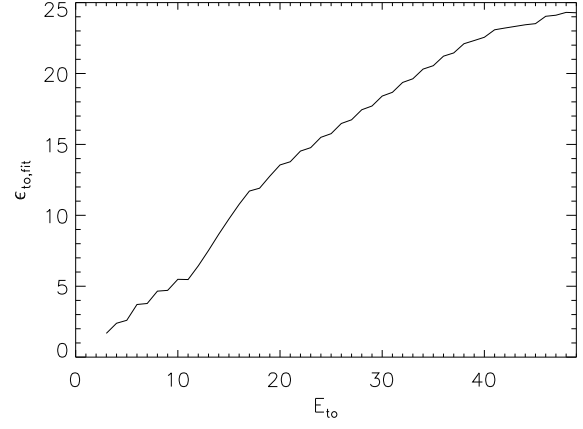}
		\caption{The photon turnover energy, $\epsilon_{to}$ (in keV), as obtained by fitting synthetic spectra with a weak $EM=0.03 \times 10^{49}$ cm$^{-3}$, $T=1$ keV thermal component,
			and a $\delta$=5, $A_{50}=10^{33}$ electrons s$^{-1}$ keV$^{-1}$ non-thermal power-law distribution with different turnover energy $E_{to}$.}
		\label{Ephto_vs_Eto}
		\end{figure}

		\begin{figure}
		\centering
		\includegraphics[width=8.8cm]{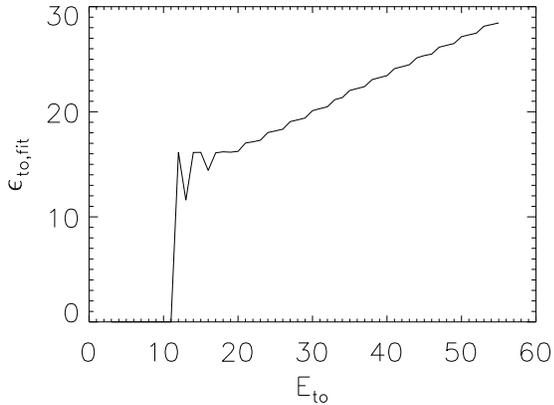}
		\caption{As in Fig. \ref{Ephto_vs_Eto}, $T=1$ keV but with a very strong emission ($EM=10^{49}$ cm$^{-3}$) for the thermal component. 
			For $E_{to} \leq$ 18 keV, the $\epsilon_{to}$ fitting parameter may take any value (below 15 keV), and still yield a very good fit.}
		\label{Ephto_vs_Eto2}
		\end{figure}

		\begin{figure}
		\centering
		\includegraphics[width=8.8cm]{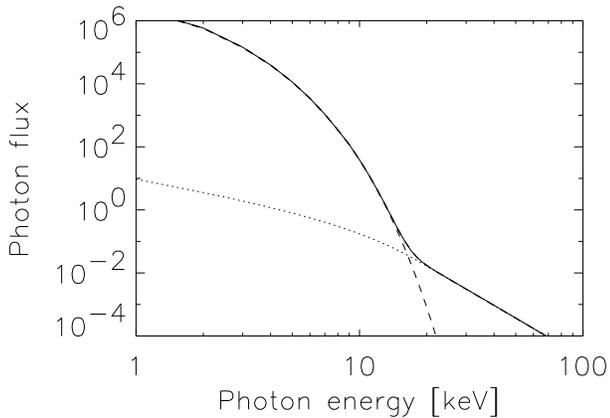}
		\caption{Synthetic spectrum obtained with $E_{to}=20$ keV, other parameters as in Fig. \ref{Ephto_vs_Eto2}. The Haug and Bethe-Bloch cross-sections have been used.}
		\label{Ephto_vs_Eto2_spectrum}
		\end{figure}

		This paragraph assumes the turnover model, but conclusions are qualitatively the same for the cutoff model.		
		As shown in Figs. \ref{Ephto_vs_Eto}, \ref{Ephto_vs_Eto2} and \ref{Ephto_vs_Eto2_spectrum}, the presence of a thermal component in the photon spectrum complicates the determination of $E_{to}$, the electron turnover energy.
		An unambiguous relationship between $\epsilon_{to}$ and $E_{to}$ cannot always be established.		
		Making several fittings with different $T$, $EM$, $\delta$, $A_{50}$, and $E_{to}$ yielded the following rule of thumb: $\epsilon_{to}$, and the $E_{to}$ derived from it,
		seem reliable only when $\epsilon_{to}>\epsilon_{th \cap nth}$ , where $\epsilon_{th \cap nth}$ is the energy where the thermal and non-thermal components of the spectrum intersect.		
		In a case such as depicted by Figs. \ref{Ephto_vs_Eto2} and \ref{Ephto_vs_Eto2_spectrum}, if the fitted $\epsilon_{to}$ is below 15 keV, then only a upper value of $\sim$18 keV may be assigned for $E_{to}$, leading to a lower boundary for the non-thermal power.
		%The situation depicted in Figs. \ref{Ephto_vs_Eto} and \ref{Ephto_vs_Eto2} are extreme cases. Most data points obtained were between those cases.

%------------------------------------------------------------------------------
\section{RHESSI flare observations, data analysis}
	\subsection{Flare selection criteria}
In the remaining part of this paper, the effects of the corrections and improvements on the computation of electron collisions and bremsstrahlung emission are studied on real data. The direction is now reversed: from the photon spectrum the electron energy distribution is derived and the total non-thermal energy is estimated. The results of the previous section serve to estimate errors.

Flares have been selected from the first 18 months of RHESSI observations, using the following selection criteria: 
		\begin{itemize}	
			\item For simplicity's sake from a data analysis point of view, flares (or portions of flares) with the same attenuator states throughout (including background time), no decimation and no pile-up were taken.
			\item Flares had to have two foot points, in order to determine a loop volume. 
				HEDC\footnote{http://www.hedc.ethz.ch} images (7" resolution, in different energy bands) were used to determine this.
			\item Significant HXR flux above 25 keV was required.
			\item Only flares above C5.0 GOES X-ray level were selected.
		\end{itemize}
From the many cases 9 flares have been selected. They were relatively simple, but some have more than one HXR $>$25 keV episode, in which case the peaks were labeled chronologically A, B, C,... All selected flares turned out to have an attenuator state of 1 (thin shutter in). Table \ref{FlaresStudied} lists them.
		%--------------
			\begin{table}				
				\centering
			\begin{tabular}{l c c c c c}
				\hline \hline
				Flare		&  	Time intervals		&pos.	\\
						&	 studied (approx.)	&	["]	\\
				\hline
				2002/04/09 12:59&	12:57:15-13:02:20	&-569,405	\\
				\hline
				2002/06/01 03:53 A&	03:53:40-03:54:35	&-423,-303	\\
				2002/06/01 03:53 ABC&	03:53:40-03:58:50	&		\\
				\hline
				2002/07/11 14:18 A&	14:17:15-14:18:10	&-791,281	\\
				2002/07/11 14:18 AB&	14:17:15-14:19:50	&		\\
				\hline
				2002/08/22 01:52 A&	01:49:25-01:50:10	&816,-272	\\
				2002/08/22 01:52 B&	01:50:10-01:55:20	&		\\
				\hline
				2002/09/08 01:39&	01:37:10-01:40:20	&-911,-205	\\
				\hline
				2002/10/05 22:50&	22:50:00-22:52:45	&-558,72	\\
				\hline
				2002/11/10 03:11&	03:07:00-03:15:00	&592,-240	\\
				\hline
				2002/11/14 11:09 A&	11:09:15-11:10:35	&-887,-262	\\
				2002/11/14 11:09 A'&	11:09:25-11:09:57	&		\\
				\hline
				2003/06/10 02:52 ABC&	02:51:15-02:53:50	&561,185	\\
				2003/06/10 02:52 B&	02:51:45-02:52:20	&		\\
				2003/06/10 02:52 C&	02:52:20-02:53:25	&		\\
				\hline
			\end{tabular}
				\caption{Flares, the time intervals that were used, and their angular offsets from Sun center.}
				\label{FlaresStudied}
			\end{table}		
		%--------------

\subsection{Extracting the thermal and non-thermal flare energies}
	
\subsubsection{Spectral fitting}
% V.1
%			Fittings were done as follow: Using the SPEX\footnote{SPEX} software package, a model composed of a thermal bremsstrahlung component and
%			a power-law component are fit to RHESSI spectra. 
%			Fitting was done usually in the 6 to 35 keV energy band for data obtained with no attenuator in or with only the thin attenuator in.
%			This insured that fitting would properly account for the thermal low-energy part.
%			For data obtained when the both attenuators were in, fitting was done in the 15-50 keV band.			
%			Using a cutoff energy at 1 keV, a best fit is found. 
%			Then, fitting were done by varying the cutoff energy in increments of 1 keV, from 6 keV to 21 keV (CHECK!!!!) (attenuator states 0 and 1),
%			or between ???? and ???? (attenuator state 3).
%			At higher values of the low-energy cutoff, the $\chi^2$ of the fit starts to grow.
%			The cutoff energy was taken to be the the low-energy cutoff at which the $\chi^2$ value of the fit was 10\% above the best $\chi^2$ value obtained.			
%			It is to be noted that in all cases studied here, having an energy cutoff at about 10 keV (as opposed to having one at 1 keV) {\it always} improved the $\chi^2$ of the fit

% V.2

		\begin{figure}
		\centering
		\includegraphics[width=8.8cm]{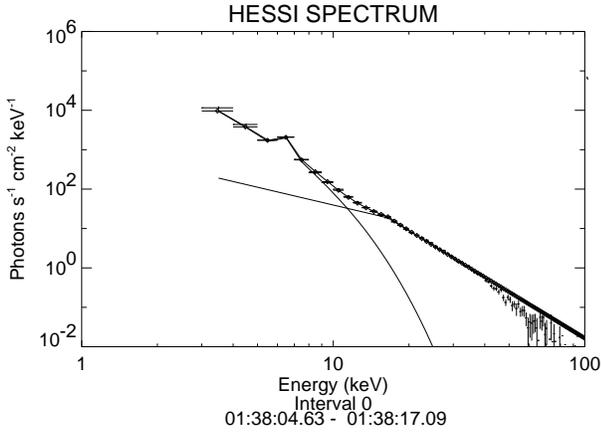}
		\caption{RHESSI photon spectrum, taken near the beginning of HXR emission of the 2002/09/08 M1.6 flare.}
		\label{spex_fit_6_35}
		\end{figure}

Using the SPEX software package of the RHESSI standard analysis tools, spectral models composed of a thermal component and a broken power-law are fitted to RHESSI spectra. 
The low-energy power-law has a fixed spectral index of 1.5. 
This somewhat arbitrary value is an average approximation of photon spectral indices at photon energies below the turnover energy.
The fitting was done for time intervals varying between 2 to 5 RHESSI spin periods ($\sim$4 s), in the 6-35 keV band. 
This band was chosen because lower energies could depend too much on the model used and the accuracy of the instrument's spectral response matrix. 
At higher energies, a spectral break may be present (e.g. Fig. \ref{spex_fit_6_35}).
A high-energy spectral break is most likely due to  a break in the original injected electron distribution (Miller 1998).
For simplicity, the details of the spectrum above this spectral break were omitted as their influence on the non-thermal power is negligible.

The five parameters retrieved are the temperature $T$ and emission measure $EM$ of the thermal component, the spectral index $\gamma$, normalization factor $F_{50}$, and the turnover photon energy $\epsilon_{to}$ of the power-law component.
Only time intervals with significant HXR flux above 25 keV were studied. 
This requirement allows better accuracy in determination of non-thermal energies. 
It also has the effect of dividing the flare into episodes of large energy input for time intervals that are short, and hence with smaller thermal energy losses.

During fitting, high values for $\epsilon_{to}$ were started with, tending to yield an upper boundary for this value in cases where it was not well-defined, as happens when the photon turnover occurs near or within the thermal part of the spectrum (which leads to a lower limit for the computed non-thermal energy).

\subsubsection{Computing non-thermal energies}
In the following, the turnover model is considered.
Every time interval yields the following fitting parameters: $\gamma$,  $F_{50}$, $\epsilon_{to}$, $T$, and $EM$. 
To compute non-thermal power or energy from Eq. (\ref{nonthermal_power_turnover}), the following must be determined: $\delta$, $A_e$, and $E_{to}$.

\paragraph{$A_e$ and $\delta$:}
To determine the correct $A_e$ and $\delta$ to use, the following method has been used, for every data point:
			\begin{itemize}
				\item	From the $\gamma$ and $F_{50}$ fitting parameters, Eqs. (\ref{Ithick2}) and (\ref{Ithick2B}) (with $A_{\epsilon}=50^{\gamma} \cdot F_{50}$), are used to determine an approximate injected electron spectral index $\delta_{approx}$ and normalization constant $A_{e,approx}$.
				\item	Using these $\delta_{approx}$ and $A_{e,approx}$, a synthetic photon spectrum using relativistic cross-sections and the albedo correction is computed.
					The resulting photon spectrum is now fitted by a power-law the same way as the observed data.
					This yields a spectral index $\gamma'$ and normalization constant $A_{\epsilon}'$.
				\item	From $\gamma'$ and $A_{\epsilon}'$, and again using Eqs. (\ref{Ithick2}) and (\ref{Ithick2B}), an injected electron spectral index $\delta'$ and normalization factor $A_e'$ are determined.
				\item	We define:
					\begin{eqnarray}
						\Delta\delta 		&=&	\delta'-\delta_{approx}	\\
						\rho_e			&=&	\frac{A_e'}{A_{e, approx}}
					\end{eqnarray}					
					If one uses the Brown (1971) method (i.e. Eqs \ref{Ithick2} and \ref{Ithick2B}) to determine the injected electron spectral characteristics from an observed photon spectrum, then the electron spectral index will have been overestimated by $\Delta\delta$ (underestimated if $\Delta\delta<0$ ), and the normalization constant by $\rho_e$ (see also Table \ref{errors_example}).
				\item	$\Delta\delta$ and $\rho_e$ depend mostly on the photon spectral index, and vary slowly with it.
					Assuming that $\Delta\delta$ is small and $\rho_e$ near unity, so that the relationship with $\gamma$ is linear, the increase in spectral index from $\delta$ (the real original injection spectral index we are looking for) to $\delta_{approx}$ is about the same as from $\delta_{approx}$ to $\delta'$, i.e. $\approx \Delta\delta$.
					Similarly, the increase in the normalization constant is similar from $A_e$ to $A_{e,approx}$ as from $A_{e,approx}$ to $A_e'$, i.e. $\rho_e$.
					Hence, the $\delta$ and $A_e$ to be used in Eq. (\ref{nonthermal_power_turnover}) can be approximated by:					
					\begin{eqnarray}
						\delta 		& \approx &	\delta_{approx}-\Delta\delta	\\
						A_e		& \approx &	\frac{A_{e,approx}}{\rho_e}
					\end{eqnarray}					

					Non-uniform target ionization effects have been neglected here: This correction seems unnecessary in light of the fact that no energy break in the relevant energy band ($<$ 35 keV) for fitting were observed.
					This could be due to the fact that expected features from non-uniform target ionization in the observed photon spectrum might go unnoticed (masking by thermal emission at the low energies, by count statistics at the high energies):
					In this case, the non-thermal energies could be overestimated by up to a factor $\sim$2.8 (Table \ref{errors_example}) for small $E_*$ (below $\sim$5 keV).
					Assuming $E_*>$25 keV, and electron spectral index $\delta > 3.5$, the error is at most $\pm$20\%, and progressively less as $E_*$ and/or $\delta$ increase.
\end{itemize}

\paragraph{$E_{to}$:}
To find the correct $E_{to}$, the following has been done:
			\begin{itemize}
				\item	Using, $\delta$ and $A_e$ as determined above, as well as the $T$ and $EM$ fitting parameters, a graph such as those presented in Figs. \ref{Ephto_vs_Eto} and \ref{Ephto_vs_Eto2} is generated.
				\item	As explained previously, one can find $E_{to}$ from $\epsilon_{to}$, or at least an upper limit for it.
			\end{itemize}

With $\delta$, $A_e$, and $E_{to}$, the non-thermal power and energy can be computed.
						
%A few of the data points in the 2002/06/01 03:53 flare were also computed using the time-consuming f\_vth\_thick.pro fitting model in the SSW/SPEX program (which assumes a low-energy {\it cutoff} in the electron distribution) as a consistency check of the above method. It yielded a total non-thermal power very close to what was derived with the present method. %, given the fact that the determination of the low-energy cutoff $E_{co}$ (the biggest source of error for the non-thermal energy) is not very reliable in both methods.

%------------------------------------------------------------thermal stuff
			\begin{table}				
				\centering
			\begin{tabular}{l c c c c c c}
				\hline \hline
				Flare		&	Class	&\multicolumn{2}{c}{Flare volume }\\
						&		&\multicolumn{2}{c}{[$\times 10^{27}$ $cm^3$]} \\
						&		&Min		&Max 		\\
				\hline
				2002/04/09 12:59&	M1.2	&	1.0	&	4.7	\\
				2002/06/01 03:53&	M1.6	&	0.48	&	3.6	\\
				2002/07/11 14:18&	M1.0	&	1.3	&	4.2	\\
				2002/08/22 01:52&	M7.8	&	1.9	&	9.6	\\
				2002/09/08 01:39&	M1.6	&	0.55	&	9.7	\\
				2002/10/05 22:50&	C6.8	&	0.12	&	4.2	\\
				2002/11/10 03:11&	M2.6	&	0.66	&	9.3	\\
				2002/11/14 11:09&	C5.9	&	0.38	&	1.7	\\
				2003/06/10 02:52&	M2.2	&	2.5	&	3.8	\\
				\hline
			\end{tabular}
				\caption{Flare volumes, as determined (Min) by RHESSI 6-8 keV imagery, or (Max) from non-thermal energies (see text).}
				\label{LoopDim}
			\end{table}		
		%--------------

		\subsubsection{Computing thermal energies}
			Thermal energies are computed using:
			\begin{equation}
				E_{th}= 3 k_BT \sqrt{EM \cdot V \cdot f}
			\end{equation}
Equal electron and ion temperatures and a unity filling factor $f$ were assumed throughout.
The assumption of a near-unity filling factor is supported by e.g. Dere (1982) or Takahashi \& Watanabe (2000).
On the other hand, there seem to be evidence that filling factors as low as 10$^{-2}$ or even 10$^{-3}$ exist (Cargill \& Klimchuk 1997), and $\sim$0.1 is an oft reported value.
Two methods were used to determine the flare volume $V$, both using RHESSI images done with the CLEAN algorithm, and with different sets of collimators (1-7, 2-7 and 3-7).
The first one consisted in making an image in the 6-8 keV band (all flares used in this study had a visible Fe-Ni line complex above the free-free continuum), estimating the flare area $A$, and using $V=A^{3/2}$ as the flare volume.
The second one consisted in looking at images made at non-thermal energies (usually 25-50 keV), estimating the size of foot points by fitting 2-D elliptical gaussians, then deconvolving for the CLEAN beam.
A volume is computed from the sizes and the distance between the foot points, assuming a perfect arc-shaped loop.
The first method yielded a lower value for $V$ and the thermal energy content, while the second one provided an upper value for both.
The range of values for $V$ easily reached an order of magnitude. 
Considering RHESSI's dynamic range, a loop 10 times larger than a smaller one may be invisible: 
Assuming both loops have the same plasma content, the surface brightness of the larger one is about $\sim$50 times lower than a that of the smaller one (cf. Appendix \ref{Appendix_sb}). 
This justifies considering the second method.
Tables \ref{LoopDim} and \ref{Ebudgets} list the flare volumes and the thermal energies derived from them.
The thermal energy {\it increases} between the start and the end of HXR flux $>$25keV were considered:
\begin{equation}
	{\Delta}E_{th}=E_{th, HXR end}-E_{th, HXR start}
\end{equation}
			
			\begin{figure}
			\centering
			\includegraphics[width=8.8cm]{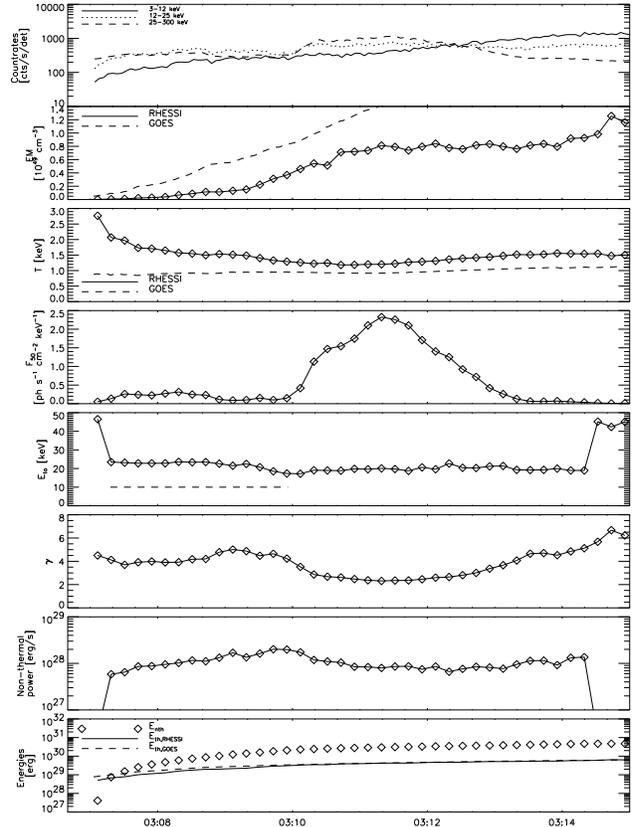}
			\caption{Data for the 2002/11/10 flare, during the main HXR peak. {\it From top to bottom}: (1) Light curves in RHESSI counts at low, intermediate, and high photon energies; (2) emission measure of thermal plasma; (3) temperature of thermal plasma; (4) calibrated photon flux at 50 keV; (5) turnover electron energy (upper limit); (6) photon spectral index; (7) non-thermal power; (8) cumulative non-thermal energy and minimal thermal energy as derived from RHESSI or GOES temperature and emission measures.}
			\label{data20021110}
			\end{figure}

%			\begin{figure}
%			\centerline{\epsfxsize=9cm \epsfbox{figs/data20020908_p1.ps}}
%			\caption{Data for the 2002/09/08 flare. Otherwise, as in Fig. \ref{data20021110}.}
%			\label{data20020908_p1}
%			\end{figure}

Figure \ref{data20021110} is an example of the acquired data and some derived quantities for the 2002/11/10 flare.
Other flares studied yielded qualitatively similar behaviours, and are not displayed here.

			\begin{table*}				
				\centering
			\begin{tabular}{l c c c c c c}
				\hline \hline
				Flare		&	Class	&	dt[s]	&$E_{nth,TO}$	& $\Delta E_{th,RHESSI,min/max}$ & $\Delta E_{th,GOES,min/max}$	\\
						&		&		&$\times 10^{30}$ $ergs$&\multicolumn{2}{c}{$\times 10^{30}$ $ergs$}	\\	
				\hline
				2002/04/09 12:59&	M1.2	&	309.4	&	2.0	&0.49/1.06	&0.48/1.04	\\
				\hline
				2002/06/01 03:53 A&	M1.6	&	98.2	&	0.40	&0.19/0.52	&0.25/0.68	\\
				2002/06/01 03:53 ABC&	M1.6	&	307.0	&	0.47	&0.58/1.59	&0.44/1.21	\\
				\hline
				2002/07/11 14:18 A&	M1.0	&	48.1	&	0.05	&0.41/0.74	&0.37/0.66	\\
				2002/07/11 14:18 AB&	M1.0	&	120.3	&	0.48	&0.48/0.86	&0.44/0.79	\\
				\hline
				2002/08/22 01:52 A&	M7.8	&	37.2	&	0.09	&0.04/0.09	&0.048/0.11	\\
				2002/08/22 01:52 B&	M7.8	&	298.6	&	21.6	&1.74/3.9	&1.53/3.4	\\
				\hline
				2002/09/08 01:39&	M1.6	&	162.0	&	6.9	&0.54/2.3	&0.46/1.9	\\
				\hline
				2002/10/05 22:50&	C6.8	&	158.7	&	0.12	&0.16/0.95	&0.21/1.24	\\
				\hline
				2002/11/10 03:11&	M2.6	&	468.1	&	4.5	&0.58/2.2	&0.55/2.1	\\
				\hline
				2002/11/14 11:09 A&	C5.9	&	72.4	&	0.20	&0.18/0.38	&0.21/0.44	\\
				2002/11/14 11:09 A'&	C5.9	&	32.2	&	2.1	&1.0/2.1	&1.1/2.3	\\
				\hline
				2003/06/10 02:52 ABC&	M2.2	&	125.9	&	6.0	&0.62/0.76	&0.48/0.59	\\
				2003/06/10 02:52 B&	M2.2	&	33.6	&	0.39	&0.19/0.23	&0.15/0.18	\\
				\hline
			\end{tabular}
				\caption{Time intervals of significant HXR flux $>$25 keV, total corrected non-thermal energy for either the {\it cutoff} or
					the {\it turnover} models, ratio of non-thermal energy over the thermal energy increase during that time interval, with the thermal energy increase computed
					using either RHESSI or GOES $T$, $EM$ data. The numbers in parenthesis are the range of values, due to the uncertainty in the thermal volume.}
				\label{Ebudgets}
			\end{table*}		
		%--------------

\section{Results and discussion}

%talking about the data plot...
As displayed in Fig. \ref{data20021110}, the peak in the 50 keV flux almost always coincides with a dip in the spectral index of the photon power-law. 
This common behaviour is thought to be a consequence of the acceleration process (Grigis \& Benz, 2004).
As observed in all our flares, the emission measure $EM$  (both RHESSI- and GOES-derived) increases during the flare.
The RHESSI-derived temperatures are usually above the GOES-derived ones, while the RHESSI-derived emission measures are below the GOES-derived ones.
However, the $T \sqrt{EM}$ product is similar most of the time (as can be deduced from the thermal energies of Table \ref{Ebudgets}).
The RHESSI-derived temperature initially decreases rapidly, then stabilizes during the rest of the time when significant HXR flux $>$ 25 keV is present.
This initial decrease in RHESSI-derived temperature is not always observed, and is never present in the GOES-derived temperatures.
The RHESSI spectral fittings were often less reliable at those early times, leading the authors to believe that the RHESSI-derived $T$ and $EM$ values are not reliable at those early times.
At the later times, thermal energies derived from both RHESSI and GOES $T$ and $EM$ yield similar values.
It might be argued that the isothermal bremsstrahlung spectrum might not always be the best model for fitting the thermal component of X-ray spectra,
and that fitting a multi-thermal model, or a full differential emission measure distribution would be more proper, although practically more difficult.

%talking about \epsilon_{to}
%It can be noted that the value of $\epsilon_{to}$, the photon turnover energy, does not vary appreciably during a HXR$>$25 keV peak (a factor $\sim$1.8 at the very most).
%Adding this $\epsilon_{to}$ break in the power-law model has proven invariably to yield better fittings than with a perfect power-law model.
%Sometimes, the $\chi^2$ fitting variable was only marginally better (down to a factor of 2 or 3), but the model with $\epsilon_{to}$ also displayed clearly better-behaved residuals.
%As previously discussed, only an upper limit on $\epsilon_{to}$ (and hence $E_{to}$) may be determined when $\epsilon_{to}$ is near or lower than $\epsilon_{th/nth}$, yielding a lower limit on the non-thermal injected electron kinetic power.
%It is also worthwhile noting that changing the upper boundary of the energy band used for fitting did not appreciably change $\epsilon_{to}$.
%More generally, the choice of a slightly different model, combined with subtle effects such as albedo or non-uniform target ionization, could certainly shift its position.

The dashed line in the fifth plot of Fig. \ref{data20021110} corresponds to times where the photon turnover energy $\epsilon_{to}$ is clearly above the thermal part, hence yielding a reliable value for the corresponding electron turnover energy $E_{to}$.
Later, the thermal part becomes so important that only an upper value for $E_{to}$ may be determined.
%The photon turnover energy, $\epsilon_{to}$, seems to anti-correlate slightly with $F_{50}$, the photon flux at 50 keV (to be investigated in a future study).
The turnover energy $E_{to}$ does not seem to change substantially during the main HXR phase, and increases to a higher value later in the flare (similar to the 2002 July 23, Holman et al. 2003).
%Moments where $\epsilon_{to}$ is larger than $\epsilon_{th \cap nth}$ occur mostly between the start of HXR emission and peak $F_{50}$.
The derived time of peak non-thermal power does not exactly coincide with the time of peak photon emission at 50 keV for this flare. 
This may not be real, as the electron turnover energy $E_{to}$ is only an upper limit at those times.

%talking about the ebudget table...
The turnover model yields non-thermal energies typically {\it only} 10 to 30 \% higher than the cutoff model.
%Fig. \ref{EphTO_Ecto_vs_delta} provides the interpretation: for a given photon turnover energy, $E_{to}$ is larger than $E_{co}$ (by typically $\sim$2 keV), leading to a smaller difference in non-thermal power.
This stems from the fact that fitting our double photon power-law on the photon spectra from a turnover electron model always yields a larger photon turnover $\epsilon_{to}$ than with a cutoff model.
This translated to a higher $E_{to}$ than $E_{co}$, leading to turnover-model non-thermal power only slightly higher than the cutoff-model non-thermal power.

The total non-thermal energy and thermal energy increase for the studied flares are summarized in Table \ref{Ebudgets}.

			\begin{table*}				
				\centering
			\begin{tabular}{c c c c c c c }
				\hline \hline
				Flare		&T,EM	&\multicolumn{5}{c}{Ratio of non-thermal to thermal energies}	\\		
				volume		&source	&All HXR peaks	&t$<$75s	&75s$<$t$<$200s	&t$>$75s	&t$>$200s	\\
				\hline
				min.		&RHESSI	&4.2$\pm$4.5	&1.5$\pm$0.9	&5.3$\pm$5.6	&5.7$\pm$5.0	&6.3$\pm$5.0	\\
						&GOES	&4.7$\pm$5.4	&1.4$\pm$1.0	&6.1$\pm$7.0	&6.5$\pm$6.0	&6.9$\pm$5.6	\\
				\hline
				max.		&RHESSI	&1.6$\pm$1.6	&0.7$\pm$0.4	&1.7$\pm$1.8	&2.1$\pm$1.9	&2.4$\pm$2.2	\\
						&GOES	&1.7$\pm$2.0	&0.7$\pm$0.4	&2.1$\pm$2.3	&2.3$\pm$2.3	&2.7$\pm$2.5	\\
				\hline
			\end{tabular}
				\caption{Ratio of non-thermal to thermal energies, for different duration $t$ of HXR emission.}
				\label{EnthEth_ratios}
			\end{table*}

Table \ref{EnthEth_ratios} lists the non-thermal to thermal energy ratios, for different durations of the HXR peak.
The non-thermal energies are lower limits, the non-thermal to thermal ratios are hence also lower limits.
Ratios obtained using the minimal flare volume are arguably closest to the truth, most notably because of filling factor considerations.
It can be noted that short HXR peaks lead to ratios of $\sim$1.5, whereas longer-duration peaks lead to higher ratios: $\sim$6.
This is expected, as radiative cooling or heat conduction (the second being most likely the dominant loss mechanism: Cargill 1994, Porter and Klimchuk 1995) tend to lower the thermal energy content, thereby increasing the ratio.
Taking time intervals ending well after the main HXR peaks, such as the SXR peak, would tend to lower the thermal energy, and to increase the non-thermal/thermal ratio.

%Emslie {\it al.} (2005), and Holman et al. (2004), using different approaches, reach similar ratios.

			\begin{figure}
			\centering
			\includegraphics[width=8.8cm]{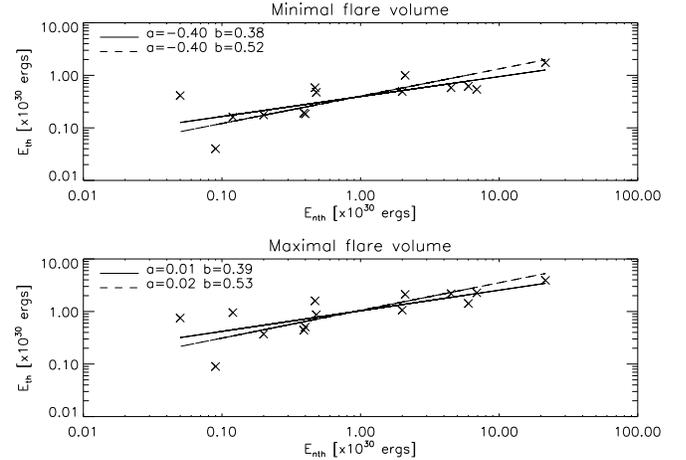}
			\caption{Log-log plot of non-thermal vs. thermal energies for all flares in Table \ref{Ebudgets}. 
				The {\it solid line} is a linear fitting, yielding constant $a$ and slope $b$.
				The {\it dashed line} is also a linear fitting, using the bisector method.}
			\label{ratios}
			\end{figure}

All data points have also been plotted on Fig. \ref{ratios}.
Linear fitting with the bisector method (Isobe et al. 1990), more relevant in cases where variables are truly independent, yields the following relation:
		
		\begin{equation}\label{EnthVSEth}
			\Delta E_{th} \sim E_{nth}^{0.5\pm0.1}
		\end{equation}

This empirical relation may simply state that the thermal energy increase $\Delta E_{th}$ does not increase as fast as the cumulative non-thermal energy, due to losses.

%The trend is to have total non-thermal energies of the same order of magnitude as the thermal energies, 
%a surprising result in light of previous studies (de Jager {\it et al.}, 1989) which usually had ratios of up to an order of magnitude.
%Of course, their instrumentation did not allow for a good determination of the low energy cutoff, and an arbitrary value had to be used.
%
%%Previously, some people used $\epsilon_{to}$ instead of $E_{to}$, if not an arbitrary 20 keV value... --> order of magnitude more non-thermal energies...
%Using $\epsilon_{to}$ instead of $E_{to}$ to compute non-thermal energies also easily yields ratios of an order of magnitude or more (Saint-Hilaire \& Benz, 2003).
%Others (Gan {\it et al.}, 2001) have used what could be the spectral break in the photon spectrum as low-energy cutoff, yielding non-thermal to thermal ratios an order of magnitude or so lower.

%------------------------------------------------------------------------------
\section{Conclusions}
The investigation of the various cross-sections approximations (relativistic vs. non-relativistic) has shown that they played a relatively minor role for our energy estimation, under usual flare conditions.
They are important, however, for spectral inversions and accurate derivation of the injected electron spectrum.
The determination of the low-energy cutoff or turnover, the largest source of error for non-thermal energies, seems reliable in some flares at the beginning of HXR emission.
With the procedure that was used in this paper, determining the non-thermal power assuming a turnover model yielded usually only slightly more ($\sim$20--25\% on the average) energy than if assuming a cutoff model.

The potentially largest sources of error for thermal energy estimations are the uncertainty in filling factors, and the fact that energy losses play a substantial role over long accumulation times.

The non-thermal energies and the thermal energies in a flare seem to be of the same order of magnitude, at least during the initial stages.
Later on, radiative cooling or heat conduction lower the thermal energy content, thereby increasing the ratio. 
The ratio of non-thermal over thermal energies given in Table \ref{EnthEth_ratios} are lower limits: Both lower low-energy cutoffs/turnovers or filling factors might increase it substantially.
A bigger than unity ratio was expected because evaporation includes heating and expansion (Benz \& Krucker 1999), and the non-thermal to thermal energy conversion might not be 100\% effective (some of the energy might rapidly be lost radiatively by cooler plasma).
%------------------------------------------------------------------------------
\begin{acknowledgements}

The authors wish to thank P. Grigis and B. Dennis for discussions.
The RHESSI work at ETH Zurich is supported, in part, by the Swiss National Science Foundation (grant nr. 20-67995.02) and ETH Z\"urich (grant TH-W1/99-2).
\end{acknowledgements}
%------------------------------------------------------------------------------
%------------------------------------------------------------------------------
%BIBLIOGRAPHY
%------------------------------------------------------------------------------

%------------------------------------------------------------------------------
\begin{appendix}

	\section{Surface brightness of two volumes with the same plasma contents}\label{Appendix_sb}
			The surface brightness $F$ of a feature is proportional to $EM/A$, where $EM$ is the emission measure ($EM=n^2V$), and $A$ the feature's area. Let $A=V^{2/3}$.
			Assuming a constant plasma content ($nV$) and temperature, $F$ is hence proportional to $V^{-5/3}$. 
			Two features possessing the same plasma content, but of differing sizes, may have a widely differing surface brightness: if the first feature is 10 times smaller in volume, it will have 46.4 times larger surface brightness!
			The total thermal X-ray flux from the smaller feature will also be 10 times larger.
\end{appendix}
%------------------------------------------------------------------------------

%\end{article}

\begin{thebibliography}{} 
\bibitem[2002]{Alexander2002} Alexander, R.C. ,\& Brown, J.C. 2002, Sol. Phys., 210(2), 407
\bibitem[2001]{Aschwanden2001} Aschwanden, M.J., \& Alexander, D. 2001, Sol. Phys., 204, 93
\bibitem[1978]{BaiRamaty1978} Bai, T., \& Ramaty, R. 1978, ApJ, 219, 705
%\bibitem[]{BenzBook} Benz, A.O., 2002: {\it Plasma Astrophysics}, 2nd Ed., Kluwer Academic Publishers
\bibitem[1999]{BenzKrucker} Benz, A.O., \& Krucker, S. 1999, 
	in Proc. 9th European Meeting on Solar Physics, 'Magnetic Fields and Solar Processes', Florence, Italy, 12-18 September 1999, ESA SP-448, December 1999
\bibitem[1971]{Brown1971} Brown, J.C. 1971, Sol. Phys., 18, 489
\bibitem[1973]{Brown1973} Brown, J.C. 1973, Sol. Phys., 28, 151
%\bibitem[1998]{Brown1998} Brown, J.C., McArthur, G.K., Barrett, R.K., McIntosh, S.W., \& Emslie, A.G. 1998, Sol. Phys., 179, 379
\bibitem[1994]{Cargill1994} Cargill, P.J. 1994, ApJ, 422, 381
\bibitem[1997]{CargillKlimchuk1997} Cargill, P.J., \& Klimchuk, J.A. 1997, ApJ, 478, 799
\bibitem[1989]{deJager1989} de Jager, C., Bruner, M.E.,Crannel, C.J., Dennis, B.R., Lemen, J.R., \& Martin, S.F. 1989, 
	in Energetic Phenomena on the Sun, Kluwer Academic Publishers, Dordrecht, Holland, p. 396
\bibitem[1982]{Dere1982} Dere, K.P. 1982, Sol. Phys., 75, 189
\bibitem[2003]{Emslie2003} Emslie, A.G. 2003, ApJ, 595, 119
%\bibitem[2005]{Emslie2005} Emslie, A.G. 2005, et al., JGR, in press
\bibitem[2001]{Gan2001} Gan W.Q., Li, Y.P., \& Chang, J. 2001, ApJ, 552, 858
%\bibitem[]{TRACE} Handy B. {\it et al.}, 1999: {\it Solar Phys.}, {\bf 187(2)}, 229
\bibitem[1997]{Haug1997} Haug, E. 1997, A\&A, 326, 417
\bibitem[2003]{Holman2003} Holman, G.D., Sui, L., Schwartz, R.A., \& Emslie, A.G. 2003, ApJ, 595(2), L97
%\bibitem[2004]{Holman2004} Holman, G.D., Dennis, B.R., \& Sui, L. 2004, AAS Meeting 204, \#54.17
%\bibitem[]{Hudson1978} Hudson, H.S., Canfield, R.C., Kane, S.R., 1978: {\it Solar Phys.}, {\bf 60}, 137
\bibitem[1990]{Isobe1990} Isobe, T., Feigelson, E.D., Akritas, M.G., Babu, G.J.: 1990, {\it Astrophys. J.} {\bf 459}, 342
\bibitem[1959]{KochMotz1959} Koch, H.W., \& Motz, J.W. 1959, Review of Modern Physics, 31(4), 920
\bibitem[2002]{Kontar2002} Kontar, E.P., Brown, J.C., \& McArthur, G.K. 2002, Sol. Phys., 210(2), 419
\bibitem[2002]{Lin2002} Lin R.P. et al. 2002, Sol. Phys., 210, 3
%\bibitem[]{LANGbook} Lang, K.R., 1998: {\it Astrophysical Formulae}, 3rd Ed., Vol. 1, Springer
\bibitem[1992]{LONGAIR} Longair, M.S. 1992, 
	in High Energy Astrophysics, 2nd Ed., Vol. 1, Cambridge University Press
\bibitem[1998]{Miller} Miller, J.A. 1998, Space Sci. Rev., 86, 79
%\bibitem[2000]{MillerRHESSI} Miller, J.A. 2000, 
	in High Energy Astrophysics - Anticipating HESSI, Astron. Soc. of the Pacific Conference Series, 206, 145
\bibitem[2004]{Petrosian2004} Petrosian, V., Liu, S. 2004, ApJ, 610, 550
\bibitem[1995]{PorterKlimchuk1995} Porter, L.J., \& Klimchuk, J.A. 1995, ApJ, 454, 499
\bibitem[2002]{PSH2002} Saint-Hilaire, P., \& Benz, A.O. 2002, Sol. Phys., 210(2), 287
\bibitem[1988]{THE} Tanberg-Hanssen, E., \& Emslie, A.G. 1988, 
	in The Physics of Solar Flares, Cambridge Astrophysics Series
\bibitem[2000]{Takahashi2000} Takahashi, M., Watanabe, T. 2000: {\it Advances in Space Research}, {\bf 25(9)}, 1833
\end{thebibliography}
\end{document}